\documentclass[11pt]{article}
\usepackage{moriond,epsfig}
\usepackage{graphicx}

\def\be{\begin{equation}}
\def\ee{\end{equation}}
\def\bea{\begin{eqnarray}}
\def\eea{\end{eqnarray}}

\begin{document}
\vspace*{3cm}
\title{Spin Correlation Effects in\\
Top Quark Pair Production\footnote{Talk presented at Rencontres de Moriond, QCD and High Energy Interactions Workshop, March 13-20, 2010 at La Thuile, Italy.}}

\author{Stephen J.  Parke  }

\address{Theoretical Physics Department\\
Fermi National Accelerator Laboratory \\
P.O. Box 500, Batavia, IL 60510, USA \\parke@fnal.gov}

\maketitle\abstracts{An analysis of the spin correlation effects in top quark pair production at hadron colliders is presented with special
emphasis for the Large Hadron Collider (LHC).  At the LHC top quark pair production is dominated by gluon-gluon fusion.  For gluon-gluon fusion at high energies the production is dominated by {\it unlike} helicity gluon fusion which has the same spin correlations as quark-antiquark annihilation.  At low energies the production is dominated by {\it like} helicity gluon fusion which imparts very strong
azimuthal correlations on the di-lepton decay products in top quark pair decay.  This production process is studied in detail and 
this suggest a new way to look for spin correlations in top quark pair production at the LHC.}

\section{Spin Correlations in Top Quark Pair Production}
At a 14 TeV proton-proton collider, the Large Hadron Collider (LHC), top quark pairs are produced 85\% of the time by gluon-gluon fusion and
15\% of the time by quark-antiquark annihilation. Thus, to understand spin correlation effects at the LHC one needs to understand
top quark pair production via gluon-gluon fusion~\cite{mp2010}.  To facilitate this understanding it is natural to separate the gluon-gluon fusion process into the contribution from {\it unlike} helicity gluons ($g_Lg_R + g_Rg_L$) and {\it like} helicity gluons ($g_Lg_L + g_Rg_R$).

For top quark pair production by {\it unlike} helicity gluons and also via quark-antiquark annihilation~\cite{mp1996}, top quark pairs
are always in an UD  or DU configuration, if the off-diagonal basis~\cite{ps1996} is used.   In this basis, the spin axis of the top quark
makes an angle, $\Omega$, with respect to the top quark direction in the ZMF.  This angle is given by
\begin{eqnarray}
\tan \Omega =  (1-\beta^2) \tan \theta,
\end{eqnarray}
where $\beta$ is the speed of the top quarks in the zero momentum frame (ZMF) and $\theta$ is the scattering angle in this frame.
Thus, at threshold, the top quark spin is aligned along the beam line, $\Omega=\theta$, and at high energy the top quark spin is aligned along the direction of the
top quark motion (helicity), $\Omega=0$; the off-diagonal basis smoothly interpolates between these two extremes. The axis for the anti-top quark
is always taken anti-parallel to the spin axis for the top quark,  see Fig 1.

CDF and D0 at the Tevatron have attempted to measure the spin correlations from quark-antiquark annihilation and have found it to be very challenging.  Since top quark pairs from {\it unlike} helicity gluon fusion have the same spin correlations as quark-antiquark annihilation, this does not seem like a fruitful way to proceed as the top quarks will be more highly boosted at the LHC.

\begin{figure*}[t]
\includegraphics{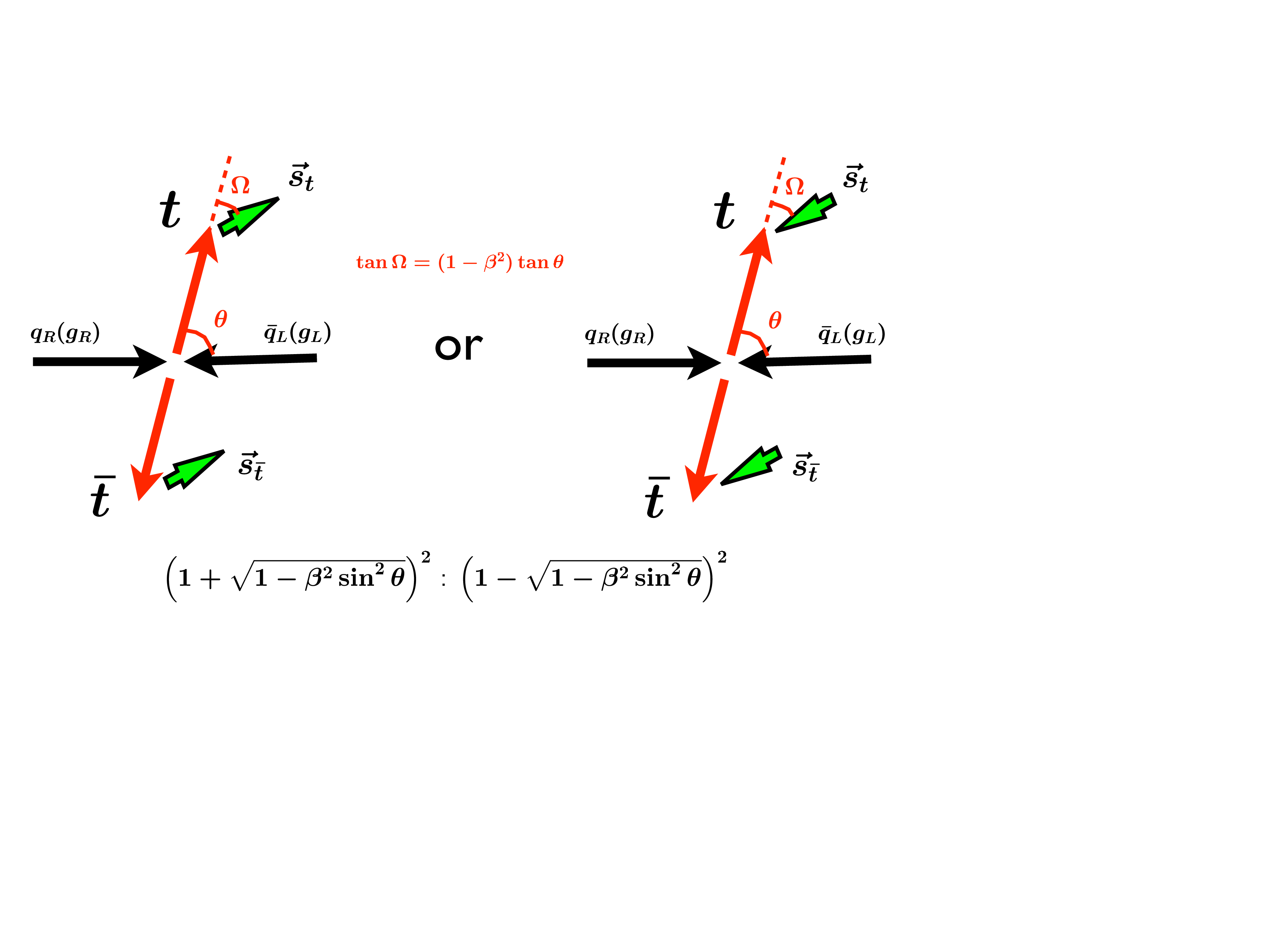}
\vspace{6.5cm}
\caption{The non-zero spin configurations for top quark pair production via quark-antiquark annihilation ($q_R\bar{q}_L \rightarrow t\bar{t}$) or {\it unlike} helicity gluon fusion ($g_R g_L \rightarrow t\bar{t}$)  in the off-diagonal basis.  This basis interpolates between the beam line at threshold, $\Omega=\theta$, and the helicity basis at ultra-relativistic energies, $\Omega=0$. For  general $\beta$, 
$\tan \Omega =  (1-\beta^2) \tan \theta$ and the relative weight of the left ($t_U\bar{t}_D$) contribution to the right ($t_D\bar{t}_U$) contribution
is given by $\left(1+\sqrt{1-\beta^2 \sin^2 \theta}\right)^2 \quad {\rm to} \quad \left(1-\sqrt{1-\beta^2 \sin^2 \theta}\right)^2$.}
\label{fig:oppo}
\end{figure*}

\begin{figure*}[t]
\includegraphics{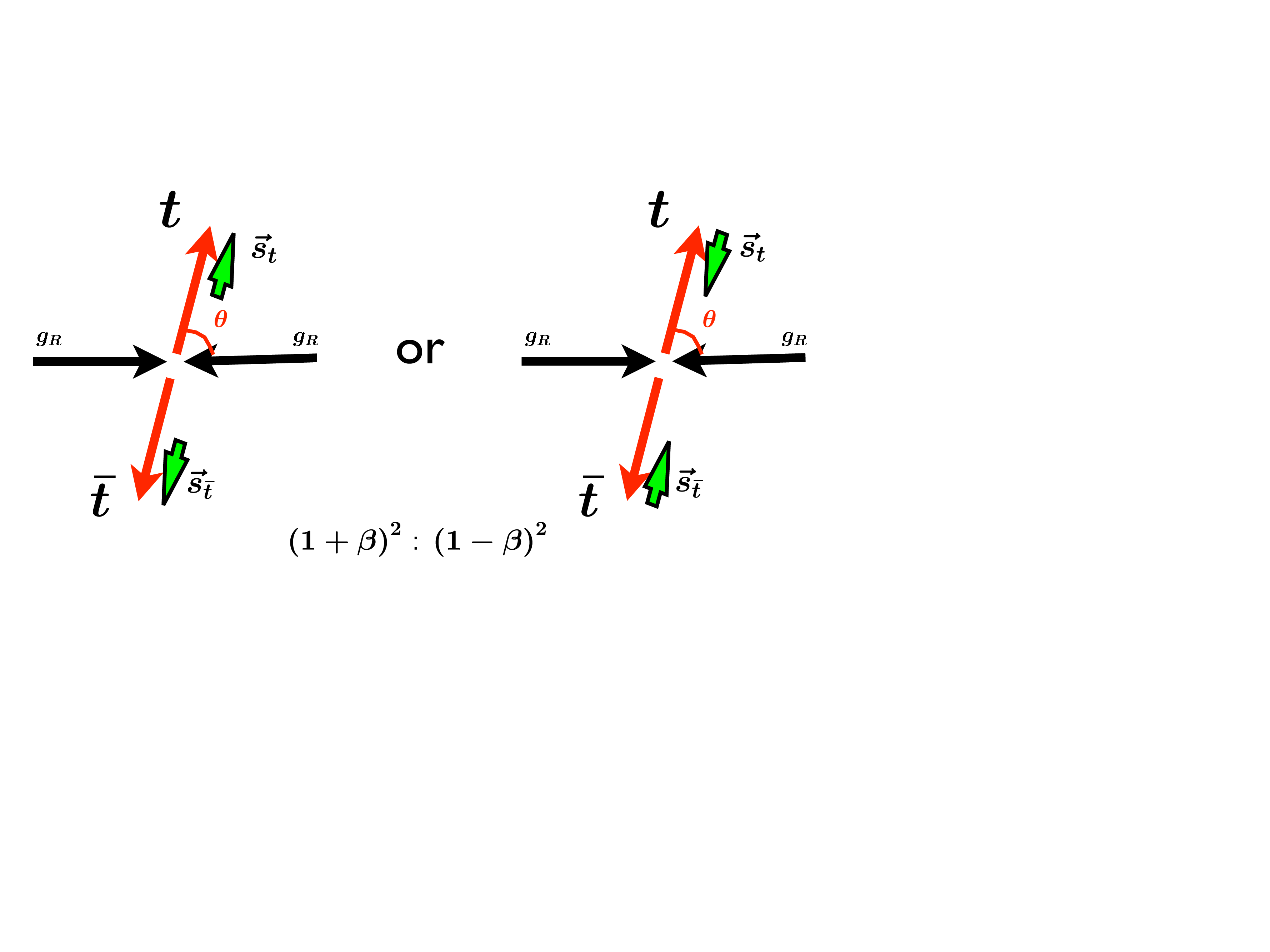}
\vspace{6.5cm}

\caption[]{The non-zero spin configurations for top quark pair production via {\it like} helicity gluon fusion ($g_R g_R \rightarrow t\bar{t}$) in the helicity basis. This is the best basis for this process.
the relative weight of the left ($t_R\bar{t}_R$)  contribution to the right ($t_L\bar{t}_L$) contribution
is given by $\left(1+\beta \right)^2 \quad {\rm to} \quad \left(1-\beta\right)^2$.
}
\label{fig:like}
\end{figure*}

So, what about top quark pairs from {\it like} helicity gluons which possibly have different spin correlations?
The contribution from {\it like} helicity gluons dominates at low invariant mass of the top quark pair whereas the {\it unlike} helicity process dominates at high invariant mass. In particular, if
\begin{eqnarray}
\beta^2< {1 \over (2-\cos^2 \theta)},
\end{eqnarray}
then the {\it like} helicity contribution dominates.  In fact the {\it like} helicity contribution to $gg \rightarrow t\bar{t}$ is 65\% of the total
contribution for top quark pair production via gluon-gluon fusion, i.e. 55\% of the total top quark pair production.

For the production of top quark pairs from {\it like} helicity gluons, the most appropriate basis is the helicity basis, where the top quark pairs are always produced in a LL or RR helicity spin state, see Fig 2.

The fully correlated, total matrix element squared for top quark production 
and decay from {\it like} helicity gluons, 
\begin{eqnarray}
g_Rg_R+g_Lg_L \quad \rightarrow \quad t+\bar{t} \quad \rightarrow \quad 
(b+\bar{e}+\nu)+(\bar{b}+\mu+\bar{\nu}),
\end{eqnarray}
is given by
\begin{eqnarray}
(\vert {\cal A}\vert^2_{RR} 
+ \vert {\cal A}\vert^2_{LL} )_{corr}
~ \sim ~
m_t^2 \{ (t\cdot \bar{e})(t\cdot \mu)
+(\bar{t}\cdot \bar{e})(\bar{t} \cdot \mu)
-m_t^2 (\bar{e} \cdot \mu)\}.
\label{eqn:like-full}
\end{eqnarray}
where, for example, $ (t\cdot \bar{e})$ is shorthand for $(p_t \cdot p_{\bar{e}})$, the dot product of the four momentum of the top quark, $p_t$, and that for the positron.  

For comparison, we define the decay of a top or anti-top quark 
into a $W$-boson and $b$-quark
uncorrelated if this decay is spherical in the top quark rest frame and 
thus independent of the top quark spin.  The $W$-boson is then 
assumed to decay in the usual (fully correlated) manner.
The uncorrelated matrix element squared is then simple given by 
\begin{eqnarray}
(\vert {\cal A}\vert^2_{RR} 
+ \vert {\cal A}\vert^2_{LL} )_{uncorr}
 \sim~ (t\cdot \bar{e})(\bar{t} \cdot \mu) 
(t\cdot \bar{t}).
\end{eqnarray}

The ratio of the correlated to uncorrelated matrix elements squared, ${\cal S}$, for like-helicity gluons 
is given by
\begin{eqnarray}
{\cal S} \equiv 
\frac{(\vert {\cal A}\vert ^2_{RR}  
+ \vert {\cal A}\vert ^2_{LL})_{corr}}
{( \vert {\cal A}\vert ^2_{RR}  
+ \vert {\cal A}\vert ^2_{LL})_{uncorr}}
 & = & \frac{m_t^2 \{ (t\cdot \bar{e})(t\cdot \mu)
+(\bar{t}\cdot \bar{e})(\bar{t} \cdot \mu)-m_t^2 
(\bar{e} \cdot \mu)\} }
{ (t\cdot \bar{e})(\bar{t} \cdot \mu)(t\cdot \bar{t}) } \nonumber \\
& = & \left( \frac{1-\beta^2}{1+\beta^2} \right)  
\left( \frac{(1+\beta^2) + (1-\beta^2)c_{\bar{e}\mu} 
- 2 \beta^2 c_{t\bar{e}} c_{\bar{t}\mu}}
{(1-\beta c_{t\bar{e}})(1-\beta c_{\bar{t}\mu})} \right),
\label{eqn:ratio}
\end{eqnarray}
where the last line is given in the ZMF in terms of  the cosine of the angles between
$t$ and $\bar{e}$ ($c_{t\bar{e}}$), $\bar{t}$ and $\mu$ 
($c_{\bar{t}\mu}$) and $\bar{e}$ and $\mu$ ($c_{\bar{e}\mu}$).
The range of ${\cal S}$ is between (2,0).  At threshold, 
$\beta \rightarrow 0$, the maximum of ${\cal S}$ occurs when the 
charged leptons are parallel, $c_{\bar{e}\mu}=+1$, whereas the 
minimum occurs when the charged leptons are back-to-back, 
$c_{\bar{e}\mu}=-1$, independent of their correlation with the 
top-antitop axis.

For non-zero $\beta$, the maximum (minimum) still occurs when the 
charged leptons are parallel (back-to-back), but they are now 
correlated with the top-antitop axis. 
The fact that the charged leptons are more likely to have their
momenta being parallel rather than back-to-back is what is 
expected for top quark pairs that have spins which are 
anti-aligned, {\it i.e.}\  LL or RR. However, here the enhancement is 
even stronger than what one would na{\"\i}vely expect because the 
interference between LL and RR strengthens the correlation between the 
momenta of the two charged leptons.  This argument suggests
looking at the 
$\Delta R$, $\Delta \eta$ and 
$\Delta \phi$ distributions of the
two charged leptons with a cut on the 
invariant mass of the top-antitop system.
However, only the $\Delta \phi$ distribution shows significant differences between the correlated and uncorrelated cases.

\begin{figure*}[t]
\includegraphics{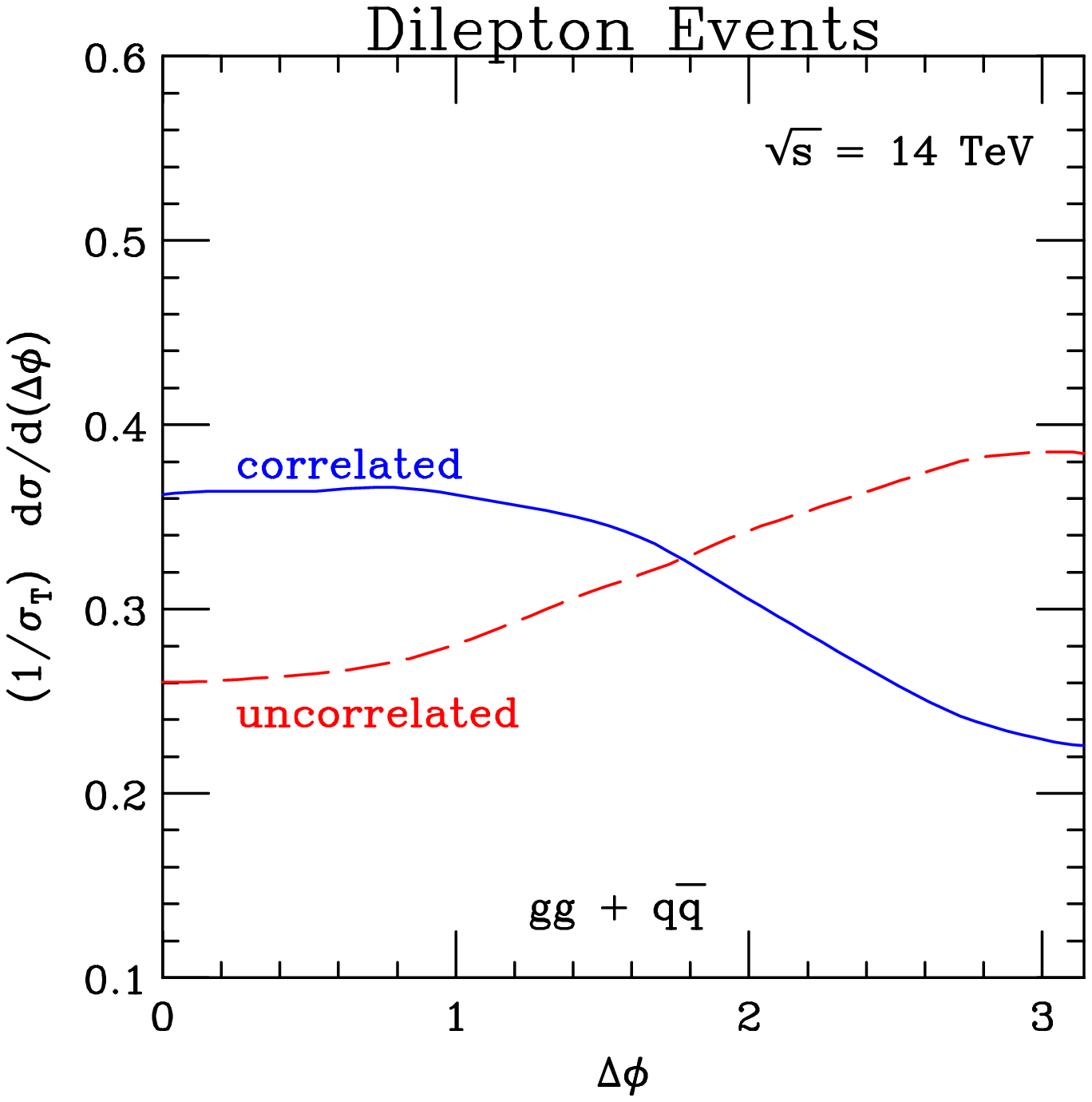}
 \includegraphics{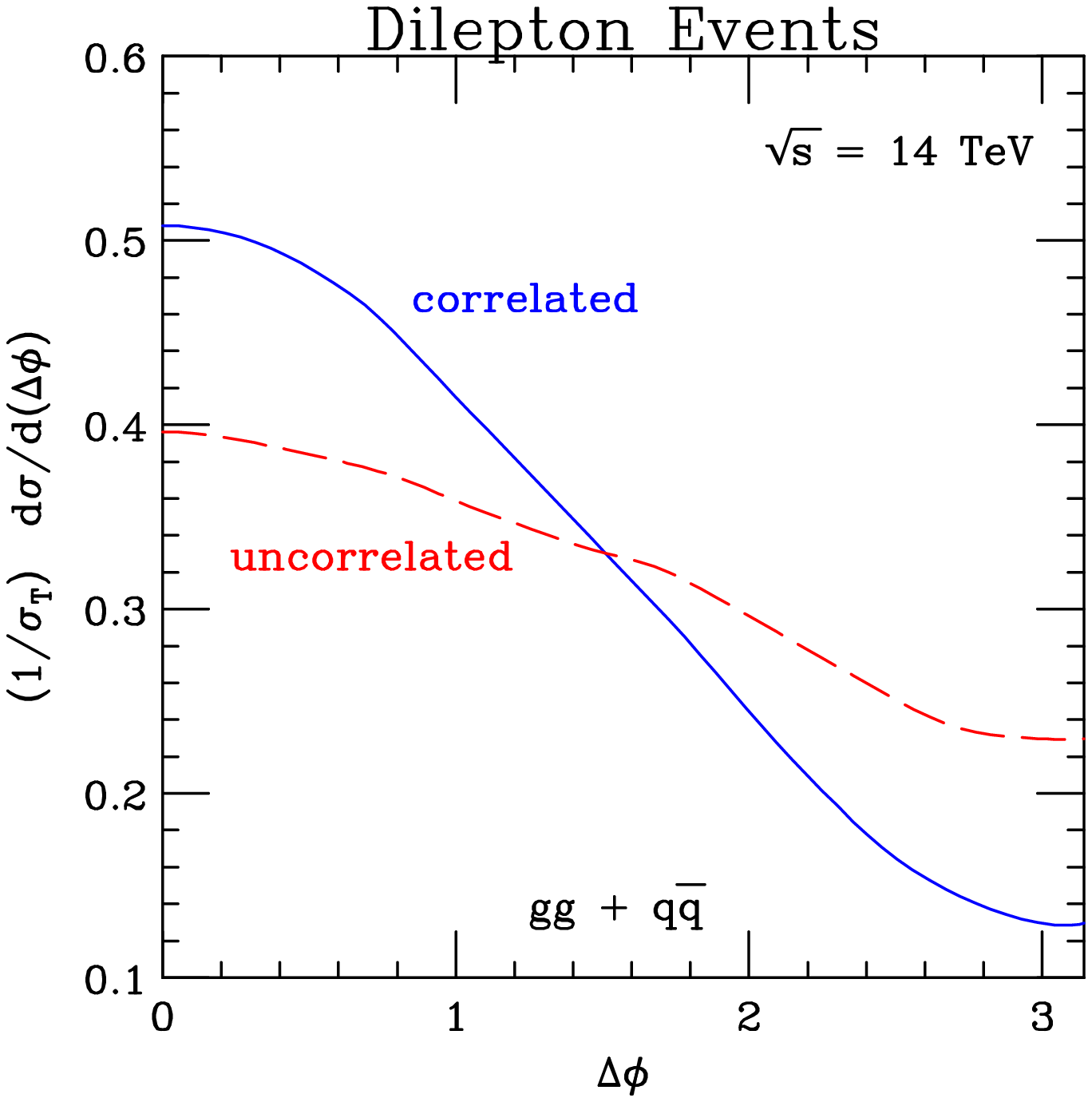}
\vspace*{7.5cm}
\caption{The differential distribution of $\Delta\phi$, 
$(1/\sigma_T)~ d\sigma/d(\Delta\phi) $.
The solid curve is for the fully correlated case
whereas the dashed curve assumes that the top quarks decay
spherically in their respective rest frames.
Left: A cut restricting the invariant mass of the $t\bar{t}$ pairs
to a maximum of 400~GeV has been applied to these distributions.
Right: A cut restricting the average reconstructed
invariant mass of the $t\bar{t}$ pairs
to a maximum of 400~GeV has been applied to these distributions. \hfill
}
\label{fig:azim-truecut}
\end{figure*}

In Fig. 3 we give the distribution $(1/\sigma_T)~ d\sigma/d(\Delta\phi) $ verses $\Delta\phi$ for the case where we have constrained  the invariant mass of the top quark pair to be less than 400 GeV (left) and where the constraint is on the average reconstructed mass
of the top quark pair (right).  Clearly, there is a difference between the correlated and uncorrelated cases using both methods of constraining the top quark pair invariant mass.  Unfortunately, the
shape of the curves changes as you go from the true invariant mass constraint to the average reconstructed mass constraint,
suggesting that this mass cut needs to be studied at NLO to properly understand these shape change effects.

In summary, a new way to investigate spin correlations in top quark production and decay is presented for the LHC.  The essences 
of this method is to look at the azimuthal angles between the two charged leptons in a sample of di-lepton quark pair events where the invariant mass of the $t\bar{t}$ system is constrained to be less than 400 GeV.  Given the large top quark pair cross section at the
LHC, it is estimated that there will be 1000 such events per  fb$^{-1}$ at 14 TeV.


\section*{Acknowledgments}
I would like to thank the organizers for a fantastic workshop and especially Vera de Sa-Varanda for making everything run so smoothly.
I would like to also acknowledge my collaborator Greg Mahlon for many important discussions on this subject. 
I, also, dedicate this proceedings to the memory of my friend and collaborator Jiro Kodaira (1951-2006) with whom I had many wonderful discussions on top quark physics as well as many other topics.
Fermilab is operated by the Fermi Research Alliance under 
contract no. DE-AC02-07CH11359 with the U.S. 
Department of Energy.

\end{document}